\documentclass[twoside]{article}
\usepackage{fleqn,espcrc2}
\usepackage{psfig}

\title{Studying  the $\omega$ properties in
pA collisions via the $\omega{\to}\pi^0\gamma$
decay\thanks{Supported by Forschungszentrum J\"ulich}}
\author{A. Sibirtsev$^1$, V. Hejny$^2$, H. Str\"oher$^2$ and
W. Cassing$^1$ \\
\vspace{3mm}
$^1$Institut f\"ur Theoretische Physik, Universit\"at Giessen,
D-35392 Giessen, Germany \\
$^2$Forschungszentrum J\"ulich, Institut f\"ur Kernphysik,
D-52425, J\"ulich, Germany}
\begin{document}
\begin{abstract}
Within transport calculations we study the production and decay of
$\omega$-mesons in $pA$ reactions at COSY energies including
elastic and inelastic ${\omega}N$ rescattering, 
the $\omega{\to}\pi^0\gamma$ Dalitz decay as well as $\pi^0 N$
rescattering. The resulting invariant $\pi^0 \gamma$ mass
distributions indicate that in-medium modifications of the
$\omega$-meson may be observed experimentally.
\end{abstract}
\maketitle

The modification of the $\omega$-meson
properties~\cite{Medium,Medium1,Medium2,Medium3,Medium4,Medium5,Medium6}
in nuclear matter have become a challenging subject in dilepton
physics from $\pi^- A$, $p A$ and $A A$ collisions. Here the
dilepton ($e^+e^-$) radiation from $\omega$'s propagating in
finite density nuclear matter is directly proportional to the
$\omega$ spectral function which becomes distorted in the medium
due to the interactions with
nucleons~\cite{Propagation,m2,m3,m4,m5,m6}. Apart from the vacuum
width (3 pion decay) these modifications are described by the real
and imaginary part of the retarded $\omega$ self energy
$\Sigma_\omega(P,X)$, where the real part $\Re \Sigma_\omega$
yields a shift of the $\omega$-meson mass pole and the additional
imaginary part $\Im \Sigma_\omega$ (half) the collisional
broadening of the vector meson in the medium. We recall that the
$\omega$-meson self energy in the $t-\rho$ approximation is
proportional to the complex forward $\omega N$ scattering
amplitude $f_\omega(P,0)$ and the nuclear density $\rho(X)$, i.e. $
\Sigma_\omega(P,X) = - 4 \pi \rho(X) f_{\omega N}(P,0)$. The
scattering amplitude itself, furthermore, obeys dispersion
relations between the real and imaginary parts
\cite{Dispersion,d2,d3} while the imaginary part can be 
determined from the total $\omega N$ cross section according to 
the optical theorem. Thus the $\omega$ meson spectral function
\begin{eqnarray} 
A_\omega(X,\!P){=}\frac{\Gamma_\omega (X,\!P)}{(P^2{-}
M_{0}^{2}{-}\Re\Sigma_\omega(X,\!P))^{2}{+}
\Gamma_\omega (X,\!P)^{2}/4} \nonumber \\  \hspace{10mm}(1) \nonumber
\label{spectral} 
\end{eqnarray} 
where $\Gamma(X,P){=}{-}2\Im\Sigma_\omega(X,P)$, can be constructed once
the $\omega N$ elastic and inelastic cross sections are known.
Note that in (\ref{spectral}) all quantities depend on space-time
$X$ and four-momentum $P$.

As mentioned before, the in-medium $\omega$-spectral function can
be measured directly by the leptonic decay $\omega{\to}l^+l^-$ or
the Dalitz decay $\omega{\to}\pi^0\gamma$ experimentally.  The
advantage of the $\omega{\to}\pi^0\gamma$ mode is due to an
isolation of the $\omega$-signal, while the dileptonic mode always
has to fight with a background from $\rho^0$ decays that are as
well produced in $pA$ collisions with compatible abundance.

The $\omega$-production in $pA$ collisions can be considered as a
natural way~\cite{Recent} to study the $\omega$-properties at
normal nuclear density under rather well controlled conditions. In
Ref. \cite{Model} the $\pi^+ \pi^-$ decays of $\rho^0$-mesons
produced in $p A$ reactions has been investigated also with the
aim of testing the in-medium $\rho^0$-spectral function at normal
nuclear matter density. However, as found in \cite{Model} the
final state interactions of the two decay pions are too strong to
allow for a decent reconstruction of the invariant mass of the
parent $\rho^0$-meson mass.  This situation changes for the
in-medium $\omega$-meson decay since here only a single pion might
rescatter whereas the photon escapes practically without
reinteraction. In this letter we thus investigate the possibility
to measure  the $\omega$-spectral function via the Dalitz decay of
$\omega$-mesons produced in $pA$ collisions.

The calculations are performed within the transport model used
before for $\rho^0$ and $K^\pm$ studies~\cite{d2,Model} by taking
into account both the direct $pN{\to}{\omega}pN$, $pN{\to}{\omega}pN\pi$  
and secondary $pN{\to}{\pi}NN$, ${\pi}N{\to}{\omega}N$ 
production mechanisms
employing the cross sections from Ref.~\cite{Elementary} as well
as ${\omega}N$ elastic and inelastic interactions in the target
nucleus (with cross sections from Ref.~\cite{OmegaN}) and
accounting for $\pi^0 N$ rescattering. Here the 
$\omega$ propagation is described by the Hamilton's equation of 
motion and its Dalitz decay to $\pi^0 \gamma$ by Monte Carlo 
according to the survival probability $P_\omega(t{-}t_0){=} 
\exp({-}\Gamma_V(t{-}t_0))$ in
the rest frame of the meson created at time $t_0$, while
$\Gamma_V$=8.4 MeV denotes the inverse $\omega$-life time in the
vacuum.

\begin{figure}[h]
\vspace{-12mm} \psfig{file=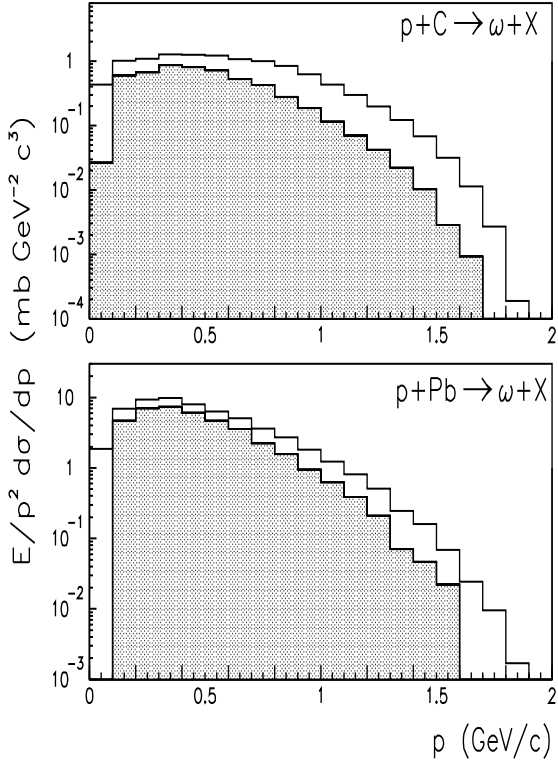,width=8.2cm,height=12.cm}
\vspace{-15mm} \caption{The invariant $\omega$-momentum spectra from 
$p{+}C$ and $p{+}Pb$ collisions at 2.4~GeV. The hatched
histograms indicate the contribution from the two-step 
($\pi N$) production
mechanism, while the solid histograms show the total yield.}
\label{ankeom5} \vspace{-4mm}
\end{figure}

Before presenting our results for the $\pi^0 \gamma$ invariant
mass distributions we show in Fig.~\ref{ankeom5} the momentum
differential $\omega$ cross section $E/p^2d\sigma_\omega/dp$ for
$p{+}C$ and $p{+}Pb$ at 2.4 GeV laboratory energy, where $p$ denotes
the $\omega$-momentum in the laboratory. The hatched
histograms indicate the contribution from the two-step production
mechanism ${\pi}N{\to}\omega{N}$, while the solid histograms
show the total yield. As can be seen from Fig.~\ref{ankeom5} the
one step mechanism $pN{\to}\omega{pN}$ dominates at large
momenta for both systems, however, the two step mechanism takes
over for low $\omega$-momenta especially in case of the heavy
$Pb$ target. These different production mechanisms, that vary
with bombarding energy and target mass, lead to a rather subtle
mass dependence of the final $\omega$-yield when parameterizing the
cross section in terms of the function $\sigma_\omega{\sim}
A^\alpha$. For a more detailed discussion we refer the reader to
Ref.~\cite{Recent}.

\begin{figure}[h]
\vspace{-15mm} \psfig{file=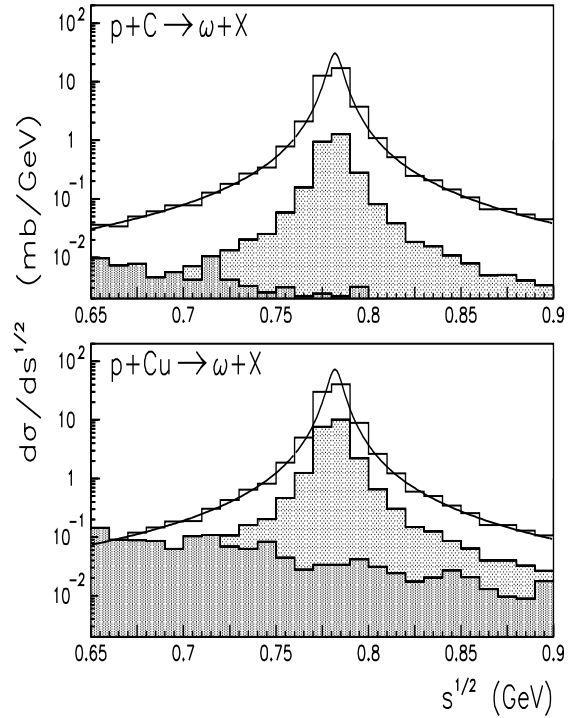,width=8.2cm,height=12.cm}
\vspace{-16mm} \caption{The $\pi^0\gamma$ invariant mass spectra from 
$p{+}C$ and $p{+}Cu$ collisions at a beam energy of 2.4~GeV
calculated without in-medium modifications of the $\omega$-meson.
The different contributions are explained in the text.}
\label{ankeom6} \vspace{-3mm}
\end{figure}

We now directly step towards the numerical results, first without
employing any medium modifications for the $\omega$-meson.
Fig.~\ref{ankeom6} shows the resulting $\pi^0\gamma$ invariant
mass spectrum for $p{+}C$ and  $p{+}Cu$ collisions at a beam energy of
2.4~GeV.  The solid histogram in Fig.~\ref{ankeom6} displays the
spectral function of $\omega$'s, which decay outside the nucleus
at densities $\rho{\le}$0.05~fm$^{-3}$; the distribution from the
$\omega{\to}\pi^0\gamma$ decay for $\rho{>}$0.05~fm$^{-3}$ for
events  without $\pi^0$-rescattering  is shown by the light
hatched areas while events that involve ${\pi^0}N$ elastic or
inelastic scattering are displayed in terms of the dark areas. The
solid lines show the Breit-Wigner distribution determined by the
pole mass and width for the free $\omega$-meson. As can be seen from 
Fig.~\ref{ankeom6} most of the $\omega$-mesons from the
$C$ target decay in the vacuum (92\%) and consequently
$\omega$ decays in the medium as well as $\pi^0$ rescattering are
rather scarce. The situation changes for a $Cu$ target where $\pi^0
\gamma$ coincidences from finite density are more frequent (19\%
$\omega$'s decay inside the target), 
however, also $\pi^0$ rescattering gives a substantial
background which in the invariant mass range of interest can
approximately be described by an exponential tail in $\sqrt{s} =
M_\omega$.

We note that experimentally the in-medium $\omega$ spectral
function can be observed by the $\pi^0\gamma$ invariant mass
distribution from $\omega$-mesons decaying inside the nucleus
without $\pi^0$-rescattering. Our results in Fig.~\ref{ankeom6}
indicate that light nuclei such as $C$ might (at first glance) not
be well suited for this purpose, however,  a $Cu$ target appears
promising.

Now we examine the feasibility of a direct detection of an
in-medium modification of the $\omega$-spectral function via the
$\pi^0\gamma$ invariant mass spectrum. The calculations are
performed for $p{+}Cu$ collisions at 2.4~GeV by introducing a real
and imaginary part of the $\omega$-potential 
\begin{equation}
\label{poten}
 U_\omega = \frac{\Re \Sigma_\omega}{2 M_0} \simeq M_0 \
\beta \frac{ \rho}{\rho_0} \end{equation} and width
\begin{equation}
\label{width} \Gamma^\ast = \frac{\Im \Sigma_\omega}{2 M_0} \simeq
\Gamma_0{+}\Gamma_{coll} \frac{\rho}{\rho_0}
\end{equation} 
in the $t{-}\rho$-approximation. Here $M_0$ and
$\Gamma_0$ are the bare mass and width of the $\omega$-meson in
vacuum while $\rho$ is the local baryon density and
$\rho_0$=0.16~fm$^{-3}$. The parameter $\beta{\simeq}$--0.16 was
adopted from the models in
Ref.~\cite{Medium,Medium1,Medium2,Medium3,Medium4,Medium5,Medium6}.
The predictions for the $\omega$-meson collisional width
$\Gamma_{coll}$ at density $\rho_0$ range from 20 to
50~MeV~\cite{Medium,Medium1,OmegaN} -- depending on the number of 
$\omega{N}$ final channels
taken into account --  and thus we examine both cases.

\begin{figure}[h]
\vspace{-14mm} \psfig{file=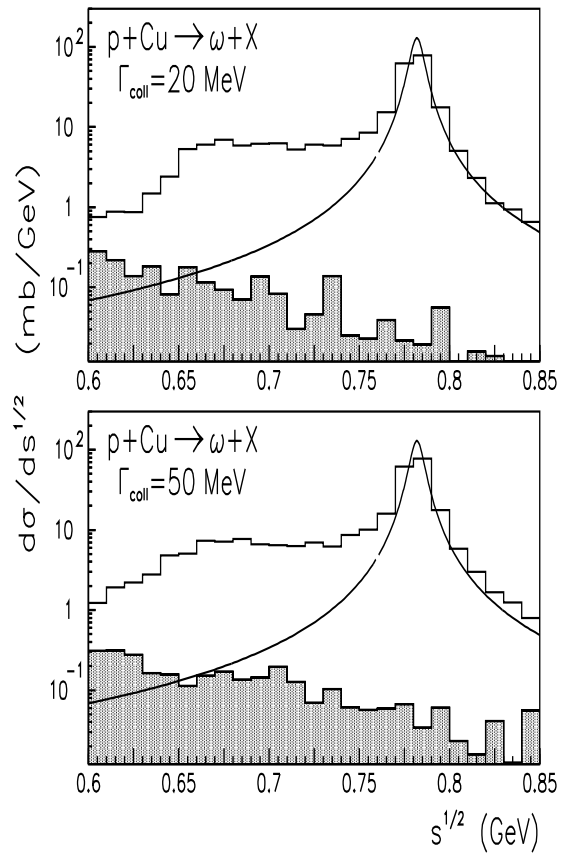,width=8.cm,height=13.4cm}
\vspace{-16mm} \caption{The $\pi^0\gamma$ invariant mass spectra from 
$p{+}Cu$ collisions at 2.4~GeV calculated for
$\Gamma_{coll}$=20 and 50~MeV (open histograms) when employing the
potential (\ref{poten}). The hatched histograms indicate the
contributions from $\omega$-mesons, that decay at finite density
and include $\pi^0$ rescattering, while the solid lines show the
$\omega$-spectral function in vacuum for comparison.}
\label{ankeom14} \vspace{-4mm}
\end{figure}

We point out that these estimates are based on the local density
approximation and neglect a momentum dependence of the
$\omega$-potential. Since the model uncertainties are quite
substantial ~\cite{Medium5,OmegaN} this situation needs
experimental clarification.

The solid histograms in Fig.~\ref{ankeom14} show the $\pi^0\gamma$
invariant mass spectra from $p{+}Cu$ collisions at 2.4~GeV
calculated with $\Gamma_{coll}$=20~MeV (upper part) and
$\Gamma_{coll}$=50~MeV (lower part) while employing the 
potential~(\ref{poten}). The results are shown for an 'experimental' mass
resolution of 10~MeV. The solid lines indicate the Breit-Wigner
distribution given by the mass and width for the vacuum
$\omega$-meson for comparison. Our results  indicate a substantial
enhancement of the low mass $\pi^0\gamma$ spectra due to the
contribution from $\omega$-mesons decaying inside the nucleus
while feeling the attractive potential~(\ref{poten}).

\begin{figure}[h]
\vspace{-6.5mm} \psfig{file=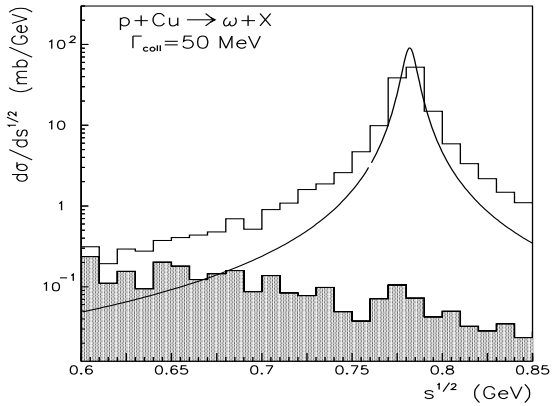,width=8.cm,height=6.3cm}
\vspace{-11mm} 
\caption{The $\pi^0\gamma$ invariant mass spectra from $p{+}Cu$ 
collisions at 2.4~GeV calculated for
$\Gamma_{coll}$=50~MeV (open histogram) when neglecting the
real part of  the
potential~(\ref{poten}). The hatched histogram indicates the
contribution from $\omega$-mesons, that decay at finite density
and include $\pi^0$ rescattering, while the solid line shows the
$\omega$-spectral function in vacuum for comparison. } 
\label{ankeom16}
\vspace{-3mm}
\end{figure}

Furthermore,
we do not observe a substantial difference between the
$\pi^0\gamma$ spectra calculated with $\Gamma_{coll}$=20 and
50~MeV since the shape of the low invariant mass spectrum is
dominated by the (density dependent) in-medium shift of the
$\omega$-pole. Only above the vacuum $\omega$ pole mass one can
see a slightly enhanced yield in case of the higher collisional
broadening. Moreover, the contribution from 'distorted'
$\omega$-mesons (due to $\pi^0$ rescattering), which is shown in
Fig.~\ref{ankeom14} by the hatched histograms, is small and
represents an approximately exponential background in the available
$\pi^0\gamma$ energy.

Now we examine the sensitivity of the predictions to the 
collisional broadening alone. The open
histogram in Fig.~\ref{ankeom16} 
shows $\pi^0\gamma$ spectrum from  $p{+}Cu$ collisions at 2.4~GeV
calculated with $\Gamma_{coll}$=50~MeV when neglecting
the mass shift of the $\omega$-meson (i.e. for $\beta$=0).
The hatched histogram again shows  the contribution from 
the 'distorted' $\omega$'s. Apparently, our calculations do 
not indicate a strong  enhancement in the low energy part of the
$\pi^0\gamma$ invariant mass spectrum, but show only the 
broadening of the distribution. 

\begin{figure}[h]
\vspace{-12mm} \psfig{file=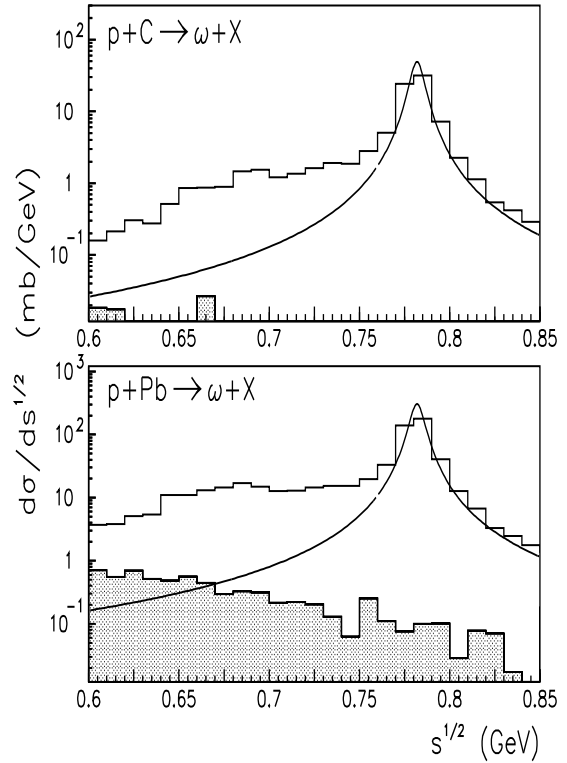,width=8.cm,height=12cm}
\vspace{-16mm} \caption{Same as in Fig.~\protect\ref{ankeom14} for
$p+C$ and $p+Pb$ collisions  at 2.4~GeV for $\Gamma_{coll}$ = 50
MeV and  $\beta$=--0.16.} \label{ankeom15} \vspace{-4mm}
\end{figure}

One might also explore further targets to optimize the signal to
background ratio. In this respect  Fig.~\ref{ankeom15} shows the
$\pi^0\gamma$ invariant mass spectra from $p{+}C$ and $p{+}Pb$
collisions at 2.4~GeV ($\Gamma_{coll}$=50 MeV, $\beta$=--0.16). 
Again there are
substantial deviations from the free $\omega$-spectral function
shown by the solid lines. Obviously the contribution from the
'distorted' $\omega$-mesons (hatched histograms) produced in $p{+}C$
collisions is almost negligible; thus light targets also might be
unambiguously used for the observation of an $\omega$-meson mass
shift  in the low mass $\pi^0\gamma$ spectrum.

In summary, we find a substantial enhancement in the low mass
$\pi^0\gamma$ invariant mass spectrum from $\omega$ decays
produced in $p{+}A$ collisions at 2.4 GeV -- relative to the $\omega$
vacuum decays --  when employing the estimates from Refs.
\cite{Medium,Medium1,Medium2,Medium3,Medium4} for the $\omega$
in-medium potential and the cross sections from Ref. \cite{OmegaN}
for $\omega N$ rescattering. These experiments might be carried
out at COSY with neutral particle  detectors looking for the 
$3\gamma$ invariant mass distribution and gating on events where 2
$\gamma$'s yield the invariant mass of a $\pi^0$. This program is
complementary to dilepton studies in these reactions,
that might be carried out with HADES detector at GSI 
Darmstadt~\cite{Hades}.
Note that in our analysis we did not consider the possible
'background' from $2\pi^0$, $\eta\pi^0$ etc. final states 
due to a finite geometry of the detector when one of the four 
photons  is out of acceptance.  

\vspace{0.5cm} The authors acknowledge valuable discussions with
E.L. Bratkovskaya, Ye.S.  Golubeva and L.A. Kondratyuk during this
study.

\end{document}